# Phase-locked feed forward stabilization for dual comb spectroscopy

Mithun Pal[1,†], Alexander Eber[1,†], Lukas Fürst[1], Emily Hruska[1], Marcus Ossiander[1,2] and Birgitta Bernhardt[1,*]

[1]Institute of Experimental Physics, Graz University of Technology, Petersgasse16, 8010Graz, Austria

[2]Harvard John A. Paulson School of Engineering and Applied Sciences, 9 Oxford Street Cambridge, Massachusetts 02138, United States

*Address correspondence to: bernhardt@tugraz.at

† These authors contributed equally to this work

**Abstract**

Sustained mutual coherence between two combs over extended periods is a prerequisite for dual-comb spectroscopy (DCS), particularly in achieving high-resolution molecular spectroscopy and precise spectral measurements. However, achieving long coherence times remains a challenge for Yb-doped frequency combs. This work introduces an experimental approach for phase-stable DCS using Yb-doped frequency combs at 1.03 µm with a novel feed-forward method, combatting the limitations of mutual coherence. Without relying on computer-based phase correction, we achieve a coherence time of 1000 seconds - three orders of magnitude longer than the current state of the art for DCS. This extended coherence enables time-domain averaging, resulting in a signal-to-noise ratio (SNR) of 2045. We demonstrate high-resolution monitoring of weak overtone transitions in the P and R branches of $C_2H_2$, with good agreement with HITRAN. The phase-locked multiheterodyne system also enables phase spectrum measurements with a scatter down to 7 mrad. Furthermore, we successfully extend our technique to the visible wavelength range using second harmonic generation, achieving high-resolution spectra of $NO_2$ with excellent SNR. The method offers high-frequency accuracy and demonstrates the potential of Yb-doped systems for multiplexed metrology, effectively extending the capabilities of DCS as a powerful tool for multi-disciplinary applications.

**Introduction**

Introduced in the late 1990s, frequency combs have rapidly become ground-breaking tools in the fields of metrology and spectroscopy. They enable the precise and absolute transfer of optical references between the optical and radio frequency domains, facilitating measurements across the entire electromagnetic spectrum. An optical frequency comb (OFC) consists of equidistant, phase-coherent spectral lines spaced by a frequency interval $f_{rep}$ and offset by $f_{CEO}$, characterized by its broad spectral range. Optical frequency combs are extensively used in many fields, including frequency metrology [1], environmental monitoring [2], medical diagnostics [3,4], material characterization [5], study of transient chemical dynamics [6], and numerous other areas of research. One of the most promising applications of OFCs in high-resolution molecular spectroscopy is their exploitation as high-brightness light sources in Fourier

Transform Spectroscopy (FTS). However, the mechanical scanning of the mirror limits the acquisition speed of high-resolution spectra with conventional scanning FTS.

Dual-comb spectroscopy (DCS) has emerged as a viable advancement to traditional FTS. This comb-enabled approach of Fourier transform interferometry eliminates the all moving parts, thereby considerably accelerating acquisition speed and spectral resolution by factors of $10^6$ and $10^3$ respectively [7]. In DCS, the beating between two asynchronous pulse trains produced by two OFCs with slightly different repetition rates creates a time-domain interferogram. This interferogram is then detected by a fast photodetector. The beating down-converts the high optical frequencies into the radio-frequency domain, which can be measured by commercial photodiodes. The scaling factor $m = f_{rep}/\delta$ is determined by the combs' repetition rates $f_{rep}$ and their difference $\delta$. The precision of DCS stems from its ability to stabilize frequency combs to known references, allowing for the accurate determination of each comb tooth's frequency ($f_n=f_{CEO}+n*f_{rep}$) using just two parameters, the carrier-envelope offset frequency ($f_{ceo}$) and $f_{rep}$. As a result, DCS achieves higher frequency resolution and accuracy compared to traditional FTS based on white light sources. DCS offers numerous benefits, including a high signal-to-noise ratio (SNR), a compact system design, and the capability for long interaction lengths due to the coherent output of the combs.

To unlock its full potential, DCS requires stabilized and phase-coherent frequency combs. Drifts or fluctuations in $f_{rep}$ or $f_{CEO}$ can reduce spectral resolution and SNR during longer averaging times. Therefore, maintaining long-term mutual coherence between the two OFCs is of crucial to ensure the technique's sensitivity. DCS has been extensively investigated using various laser platforms across a range of spectral windows. In the near-infrared (NIR) wavelength range, mode-locked erbium-doped fiber lasers [8–10], Ti:Sa lasers[5], and frequency combs generated by the strong modulation of a continuous-wave (CW) laser with electro-optic modulators (EOMs) [11,12] are the predominant technologies due to their mature optics. The majority of phase-stabilized DCS methods have been demonstrated using these laser sources, which benefit from the high-bandwidth feedback electronics. Ytterbium fiber frequency combs centered at 1 μm are less commonly utilized in DCS. However, they were employed in the early demonstrations of cavity-enhanced DCS for investigating molecules with low absorption cross sections [13]. Recent demonstrations of visible (VIS) [14] and ultraviolet (UV) DCS [15] using mode-locked ytterbium fiber-based combs to probe ro-vibronic transitions of important atmospheric trace molecules have highlighted the potential of these lasers. However, the stabilization of the $f_{ceo}$ and maintaining coherence between two comb sources are constrained by the limited degrees of freedom, primarily due to the low bandwidth electronics associated with the pump diodes.

A conventional approach to phase-stabilize OFCs employs the *f*-2*f* scheme with an octave-wide spectrum [16,17]. However, this method suffers from its complexity and sensitivity to intensity fluctuations. An alternative approach to maintain the coherence of two OFC is by phase-locking both combs to the same pair of cavity-stabilized CW lasers. This is achieved by utilizing fast and slow feedback loops and enables coherent averaging time of 1 second [9]. Chen et al. demonstrated the feed-forward stabilization of the relative carrier-envelope offset frequency between two femtosecond erbium-doped amplified fiber lasers using an acousto-optic frequency shifter (AOFS) [18]. This approach achieved an experimental coherence time of 2000 seconds without indications of saturation. To achieve even longer averaging times, real-time adaptive sampling [19] and computational correction techniques [20] have been employed. By using auxiliary CW narrow line-width lasers, these methods permit continuous adjustments and corrections of phase slippage during data acquisition, significantly enhancing the coherence time.

Furthermore, the mutual coherence of the combs can be reconstructed using self-referencing correction numerical algorithms from two free-running frequency combs [24]. These algorithms correct for phase errors without the need for supplementary measurements or additional optical elements, making the system more robust and easier to manage. However, this approach introduces a significant challenge in terms of data storage and processing, and their effectiveness is limited by the time between two consecutive dual-comb interferograms [21]. Nowadays, single-cavity dual-comb sources, such as thin-disk oscillators [22] and polarization multiplexing in MIXSELs [23], are becoming popular for coherent dual-comb spectroscopy due to their pulse trains sharing the same cavity components and noise characteristics, resulting in high mutual stability.

In this article, we introduce a novel and straightforward phase-locking technique for achieving coherent averaging in a NIR dual-comb spectrometer. The spectrometer employs two broadband ytterbium-doped fiber frequency combs around 1 μm. By beating both combs with a single free-running CW laser, we track their relative carrier-envelope offset frequency $\mathit{\Delta}f_{ceo}$ and actively compensate for the fluctuations between both lasers using an AOFS. In contrast to existing feed-forward schemes -which directly amplify the detected $\mathit{\Delta}f_{ceo}$ and feed it to an AOFS - we drive the AOFS using the output of a high-power voltage-controlled oscillator (VCO) that is phase-locked to $\mathit{\Delta}f_{ceo}$. This scheme effectively isolates our comb output power from any intensity fluctuations in the weak $f_{ceo}$ detection signals. Subsequently, it bypasses the bandwidth limitations typically encountered in Yb-doped fiber laser systems' electronics, which can restrict the effective response time for phase stabilization.

We experimentally demonstrate a mutual coherence time exceeding 1000 seconds, which is three orders of magnitude higher than previously reported for other ytterbium-coupled DCS systems for coherent averaging. This extended coherence time enabled us to record a 2-second-long-time trace, generating comb-resolved spectra that easily resolved individual comb modes spaced at 80 MHz, with a full-width at half-maximum (FWHM) of 1.3 MHz for a single comb mode. In the time domain, a SNR of 34,000 was achieved by averaging over 1000 interferograms for a time-averaged period of 42 ms. Furthermore, we employed this coherent DCS system to observe weak overtone transitions in acetylene ($C_2H_2$), demonstrating its capability for high-resolution spectroscopic analysis of trace gases. The system's long-term stability and mutual coherence pave the way for conducting coherent DCS experiments in the VIS and UV wavelength regions via high harmonic generation, enabling sensitive measurements of atmospheric trace molecules.

**Experimental Design**

Figure 1 illustrates the experimental set-up of our near-infrared dual-comb interferometer. Two commercial mode-locked ytterbium-doped fiber lasers (Menlo Systems GmbH orange Femtosecond Ytterbium Laser), emitting over a bandwidth of 8.4 THz (30 nm) centered around 289 THz (central wavelength 1036 nm). The lasers operate with repetition rates of 80 MHz and 80 MHz + 24 Hz. The repetition rates of both combs are stabilized against a common 10 MHz reference using a feedback loop controlled by piezo actuators, which adjust the cavity lengths to actively compensate for any fluctuations in $f_{rep}$. An AOFS (Brimrose TEF-120-80-1030) is placed in the beam path of one of the lasers (Laser B in Figure 1), diffracting a portion of the laser output into the first diffraction order. The spectrum of this diffraction order is frequency-shifted from the original comb by the frequency of the radio frequency (RF) wave driving the AOFS, $f_{AOFS}$, and its power depends on the amplitude of the RF wave. The frequency-shifted first-order beam is spatially overlapped with the second laser (Laser A in Figure 1) using a beam combiner, generating periodic bursts in the time domain known as interferograms. These interferograms occur every $1/\delta$, where $\delta$ is the difference between the repetition rates of the lasers.

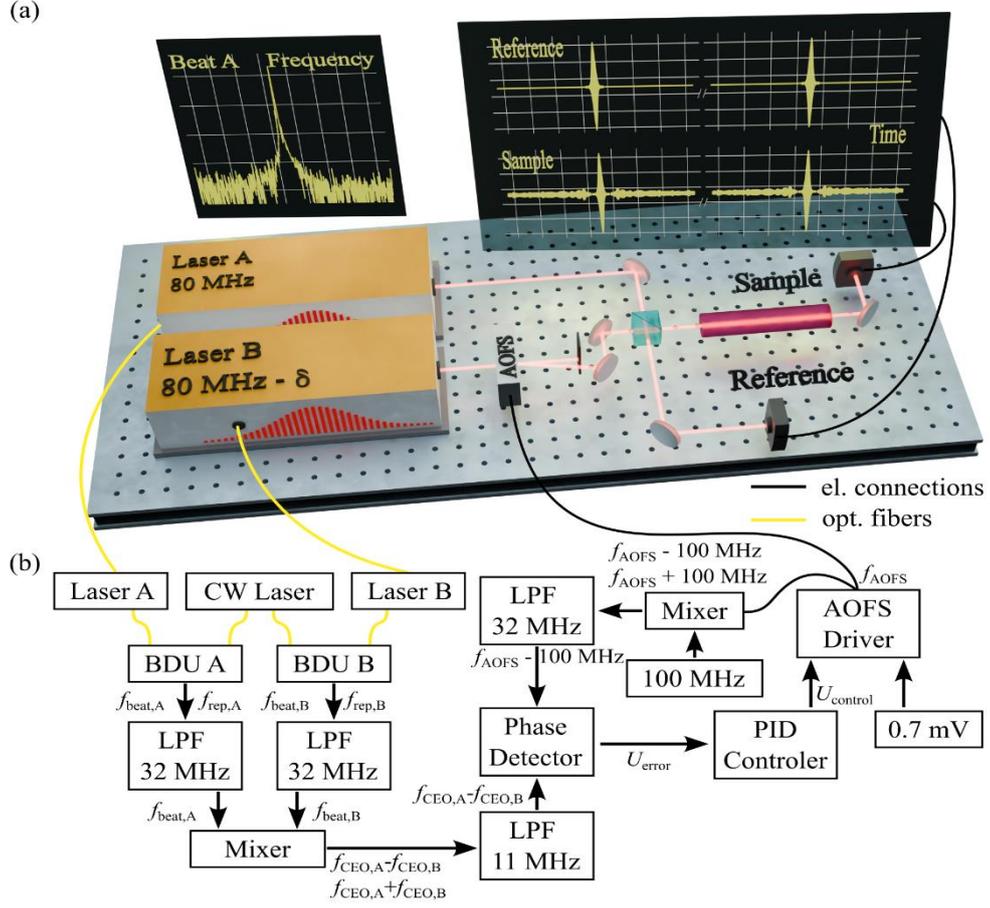

*Fig. 1. Schematic of the used experimental set-up to record phase stabilized absorption spectra.* (A) schematic of the optical set-up of the phase-locked dual comb spectrometer. Laser A is operated at a locked repetition rate of 80 MHz, while laser B is stabilized at 80 MHz - $\delta$ with a detuning of $\delta$ = 24 Hz. Laser B is passed through the AOFS and the first order is superimposed with the output of laser A on a beam splitter cube. One output of the beam combiner is sent through the sample while the second is used as the reference. (B) schematic of the electronic circuitry used to stabilize the carrier envelope offset frequency difference between the two combs $\Delta f_{ceo}$. The frequency combs generate individual beat signals in two beat detection units (BDU), which are subjected to low-pass filtering, mixing, additional filtering, and subsequent comparison to a reference signal. This reference signal is generated by mixing down a monitoring signal of the AOFS driver output with a stable 100 MHz signal. The resulting phase change is used as the input to a PID controller, closing the phase-locked loop to the AOFS driver.

The interferograms are detected simultaneously by two high-speed photodetectors (Thorlabs PDA10A2). One detector measures the signal after interaction with the sample, while the other provides a reference by measuring the beam without the sample, ensuring sample-free detection for comparison. The detector signal is then passed through a custom band-pass anti-aliasing filter (3-30 MHz) before being fed into a 16-bit, 250 MSa/s digitizer card for data acquisition.To enable coherent averaging in the time domain, we stabilize the relative carrier-envelope offset frequency between the two combs using a feed-forward scheme [18]. The $\Delta f_{ceo}$ between the two frequency combs is measured by interfering each comb individually with a free-running, narrow-linewidth (<20 kHz) CW laser (NKT Photonics Koheras Adjustik). The lowest order radio frequency beat notes $f_{beat, A/B} = f_{n, A/B} - f_{CW} = f_{CEO, A/B} + n_{A/B}*f_{rep, A/B} - f_{CW}$, are generated by the beating between the CW laser and the nearest comb line of each optical frequency comb (OFC). These beat notes are detected by a low-noise IR detector (Qubig PD-100-SWIR1), achieving SNR greater than 35 dB. A low-pass filter (Mini-Circuits BLP+36+) removes high-frequency components (e.g., higher-order beat notes and $f_{rep}$ before the signal is amplified to +5 dBm using a +30 dB electronic amplifier (Mini-Circuits ZFL-500LN). To ensure reliable detection of both beat notes, $f_{beat,A/B}$, and their difference, we position the beat frequencies at approximately 15 MHz and 25 MHz using waveplates in the fiber combs.

Subsequently, the two RF beat notes are electronically mixed (Mini-Circuits ZAD-1-1+), producing both the sum and difference frequencies, $f_{beat,A} \pm f_{beat,B}$. The sum frequency is discarded using a low-pass filter, leaving the difference signal, $\Delta f_{beat} = f_{beat,A} - f_{beat,B} = f_{CEO,A} + n_A*f_{rep,A} - f_{CEO,B} - n_B*f_{rep,B}$. In this difference signal, the CW laser frequency cancels out (along with its fluctuations), allowing us to monitor the $\Delta f_{ceo}$ variations.

Meanwhile, $f_{CEO,A}$ of Laser A is left fluctuate freely without any active control or stabilization. The jitter in the $\Delta f_{CEO}$ (= $f_{CEO,A} - f_{CEO,B}$) is eliminated by frequency down-shifting the first diffraction order generated from the output of Laser B by the AOFS, using $f_{AOFS} = \Delta f_{beat} + f_{const}$ regardless of the constant frequency offset $f_{const}$. In conventional feed-forward methods, $\Delta f_{beat}$ is directly amplified to drive the AOFS, with $f_{const}$ applied to align it within the AOFS's efficiency range. However, the diffraction efficiency of the AOFS depends on the driving power, meaning that fluctuations in the beat note signals can propagate to the output power of the stabilized laser. To mitigate this, we drive the AOFS using a 33 dBm voltage-controlled oscillator (VCO), where both the frequency and power are independently controlled by adjusting two DC voltage inputs. This approach ensures that the first-order diffracted beam and the interferogram are immune to intensity fluctuations, resulting in a reduction of relative intensity variations of the pulse train to below 0.5 %. These fluctuations were measured using a photodiode and calculated based on the method outlined by Smith et al., 2022 [27].

To lock the frequency of the voltage-controlled oscillator to fluctuations in $\Delta f_{CEO}$, we redirect a small fraction of its output ($f_{AOFS}$) to a phase detector (MenloSystems GmbH DXD200) which compares it to the $\Delta f_{beat}$ signal. The phase detector measures the phase difference (error) between two signals and converts this into a voltage signal. This voltage signal is sent to a commercial proportional-integral servo (Newfocus LB1005), whose output controls the VCO's frequency modulation (FM) input, forming a phase-locked loop and allowing for frequency control with a bandwidth of up to 10 MHz. Combined with the AOFS's rapid response time (60 ns), this method overcomes the bandwidth limitations of the pump diode's current modulation, enabling a fast-acting stabilization. A technical detail worth mentioning is that we use a mixer to frequency down-shift the redirected portion of the VCO's output signal (oscillating at $f_{AOFS}$) by a constant offset $f_{const}$ = 100 MHz before it reaches the phase detector. This down-shift allows us to drive the AOFS in its optimal efficiency range (around 110 MHz) while still locking it to $\Delta f_{ceo}$ (around 10 MHz).

**Results and discussion**

To demonstrate the system's performance, we investigate the absorption spectrum of acetylene ($C_2H_2$), a spectroscopic prototype molecule. Its rather weak overtone absorption band at around 289 THz (1.03 µm) is perfectly suited to qualify the capability of the phase locked loop feed forward stabilization method introduced here. We employ a 35 cm-long gas cell in a triple-pass-configuration, providing an effective absorption path length of 105 cm. The cell is equipped with Brewster windows to minimize reflections and etalon effects and it is positioned in the combined beam path of the two combs. Prior to measurements, the cell is flushed with argon, evacuated, and then filled with (200 ± 10) mbar acetylene at natural abundance. To avoid detection non-linearities or artifacts in the photodiode signal, the optical power of the combined beam is kept below 1 mW.

Figure 2(a) displays a 170 ms-long trace featuring five consecutive interferogram bursts, recorded under phase-locked conditions. The repetition rate difference ($\Delta f_{rep}$ = 24 Hz) corresponds to the observed temporal separation between bursts of $1/\Delta f_{rep}$ = 41.6 ms. The data was sampled at 250 MSa/s, yielding a total of 42.5 million data points. However, the benefits of time-domain coherent averaging for long traces with low detuning are limited by the memory usage constraints of the PC. Therefore, a single-shot interferogram (IGM) was recorded across a 41.6 ms time span, followed by continuous time-domain coherent averaging. The primary advantage of this averaging approach is the significant reduction in data file size, which circumvents memory limitations and accelerates computation.

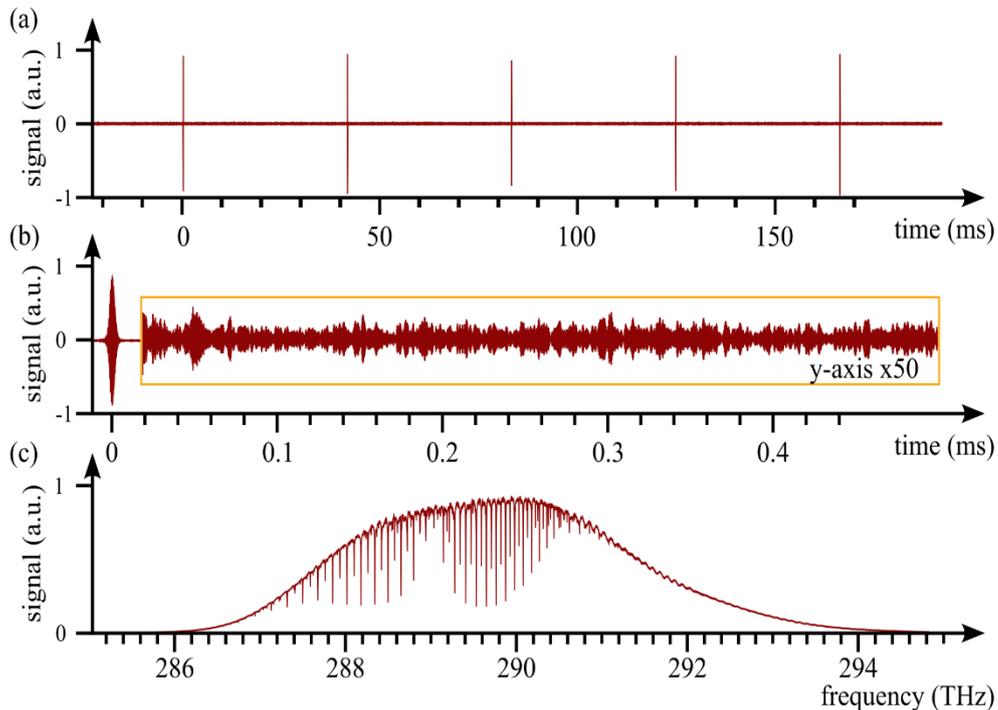

*Fig. 2. The molecular response of acetylene in the NIR. (A) interference pattern train spaced by 42 ms, corresponding to the detuning of 24 Hz. (B) one-sided temporal free induction decay response of acetylene, averaged over 2500 interference patterns using a 42-ms long apodization window. (C) the transmission spectrum of acetylene obtained via Fourier transforming the data in panel b and frequency upconverting into the optical domain.*

Figure 2(b) illustrates an example of an individual IGM, featuring a central burst caused by the optical delay of $1/f_{rep}$ = 12.5 ns between two consecutive pulses, followed by a trailing electric field indicative of molecular free induction decay (FID). This decay reflects the gradual emission of radiation as excited molecular states return to equilibrium. Real-time coherent averaging is achieved by simply co-adding 2500 IGMs for 100 seconds total measurement time. The signal is recorded with a time span of $1/\Delta f_{rep} = 41.6$ ms to attain the maximum spectral resolution of 80 MHz. This coherent averaging significantly enhances the SNR in the time-domain signal, allowing clear observation of such subtle molecular FID features. Without coherent averaging, this detail is completely obscured by noise.

A Fourier transform of the averaged interferogram and scaling by the down-conversion factor ($\Delta f_{rep} / f_{rep} = 3 \times 10^{-7}$) reveals the transmissive optical spectrum of $C_2H_2$, shown in Figure 2(c), over a spectral span of 8 THz. Coherent averaging allows for the clear identification of the P and R transitions from weak overtone bands ($3\nu_3$) and combination bands ($2\nu_1 + \nu_3$, $3\nu_3 + \nu_4 - \nu_4$, $\nu_1 + \nu_2 + \nu_3 + 2\nu_4$) of $C_2H_2$. An etalon effect, likely caused by spurious reflections in the transmissive optics, is also observed in the spectra. However, this issue is resolved by monitoring the reference IGM with a similar detector placed in the other arm of the setup.

Next, to evaluate the spectral performance of the current DCS setup we capture multiple consecutive interference patterns over extended measurement times, which enables the resolution of individual optical frequency comb modes. To manage data size, we increase the repetition rate difference to $\Delta f_{rep} = 50$ Hz, which decreases the temporal spacing between consecutive interferograms to 20 ms. Additionally, the acetylene pressure is reduced to 65 mbar to mitigate pressure broadening effects. Figure 3(a) shows the comb-resolved spectrum of the $C_2H_2$ ro-vibrational transitions obtained by Fourier-transforming a single, 2-second long, measurement including 100 sequential interferograms.

The expanded view of a comb-resolved $C_2H_2$ ro-vibrational transition is shown in Figure 3(b), which clearly resolves the Doppler FWHM of 349 MHz at 298 K. It also highlights the effectiveness of coherent averaging, as the individual comb lines, spaced at 80 MHz, precisely probe the molecular transition, achieving high resolution. This level of resolution is sufficient to sample most molecular profiles in the infrared spectral region at room temperature, offering detailed insight into the molecular transitions. A total of 85,000 distinct comb lines are resolved across the entire spectral range. For an averaging time of 2s and a detuning of 50Hz, the FWHM of each comb line is measured to be approximately $(0.80 \pm 0.05)$ Hz in the radio frequency domain, as determined by a Gaussian fit. This corresponds to $(1.28 \pm 0.08)$ MHz in the optical domain, which aligns well with the Fourier transform limit of 0.8 MHz. The maximum SNR of individual comb lines is achieved at 300 at the center of the optical spectrum. Such a robust SNR enhances the accuracy and quality of the molecular transitions being probed, allowing for precise spectral measurements.

Comb modes are resolved even at the spectral wings, though the decreased power per mode reduces the SNR. In the absence of a locking scheme, the comb modes broaden to the extent where they cannot be resolved anymore. However, it should be noted that high-detuning compromises the spectral bandwidth in this scenario.

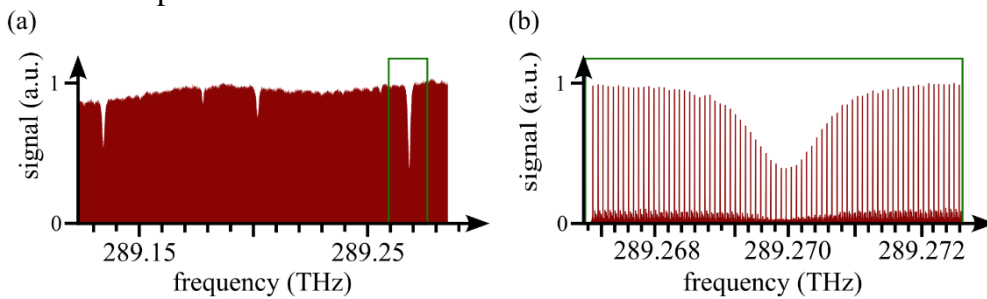

*Fig. 3. Comb-resolved spectrum of acteylene. Data was retrieved from a single, 2-s long, measurement, using 50 Hz detuning (containing 100 interference patterns) (A) resolved heterodyne beat signals between optical frequency combs, exhibiting molecular absorption dips. (B) a closer view showing comb-tooth-resolved optical absorption spectra with the resolved comb modes spaced by the repetition rate of the laser, 80 MHz.*

Far away from the CW laser wavelength (~289 THz), we observe the extra comb lines emerging This can be attributed as phase fluctuations at the CW laser wavelength are fully compensated, while some phase fluctuation further away remain.

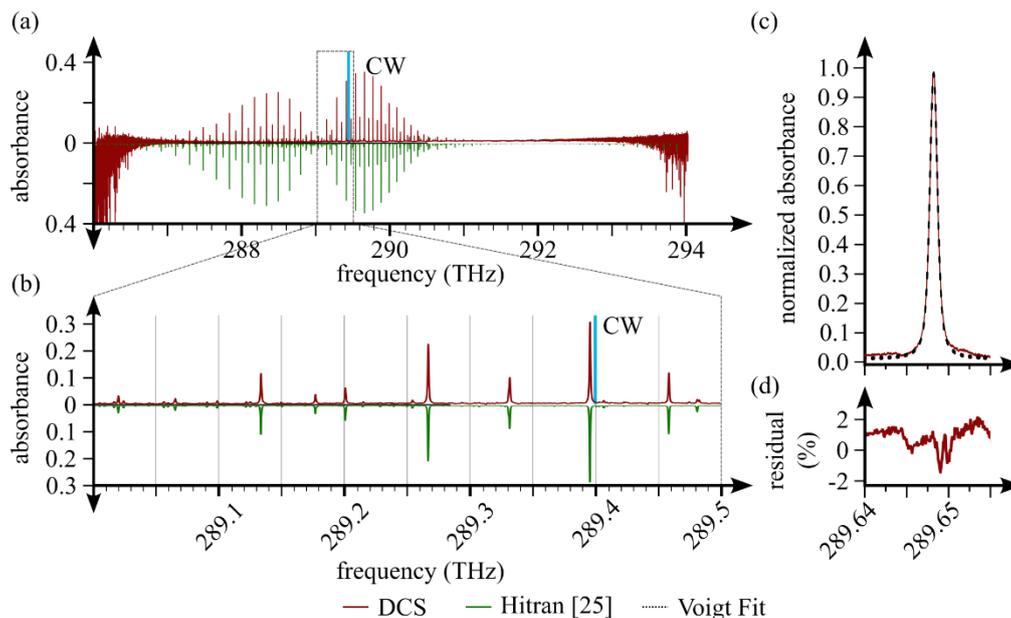

*Fig. 4. Comparison of acetylene absorption with literature. (A) the $C_2H_2$ absorption spectra from our locked DCS (red line) compared with the HITRAN database (mirrored green line). (B) a zoom into the region close to the CW laser (indicated by the blue vertical line) to facilitate the comparison of the line positions and line strengths. (C) a zoom into the highest peak in the right absorption band with a Voigt fit (black dotted line). (D) the residual between the measurement data and the Voigt fit in Figure 4(c)*

We validate the spectral quality obtained from our dual-comb spectroscopy (DCS) with the HITRAN database [25]. Figure 4(a) compares the absorbance spectrum of $C_2H_2$, obtained from 5000 averaged interferograms, with a spectrum calculated from the HITRAN database under the same experimental conditions (pressure: 65 mbar, path length: 105 cm, temperature: 298 K). The absorbance was calculated using Beer-Lambert law ($A=\log(I_0/I_t)$, $I_0$: incident intensity, and $I_t$: transmitted intensity). An additional baseline correction routine, based on a polynomial fit, was applied to remove local baseline variations caused by small discrepancies between the two photodiodes, further improving the accuracy of the measurement. The comparison highlights the excellent performance of our phase-locked dual-comb spectrometer. The high signal-to-noise ratio in the phase-locked averaged spectra clearly reveals the P-branch and R-branch rotational lines of the $3\nu_3$ overtone band, which originate from the symmetric stretching mode of the $C_2H_2$ molecule. The mean relative uncertainty between the phase-locked DCS spectrum and the HITRAN-computed spectrum in absorbance is 10.63%. This uncertainty increases as we move further away from the CW laser frequency. To determine the line positions in our self-calibrated spectra, we fitted Voigt profiles to the experimental absorption data. This fitting also allowed us to assess the contribution of instrumental broadening by comparing the experimental data with the computed spectra, which showed minimal impact due to coherent averaging over an extended period close to the CW laser wavelength. Figure 4(c) and (d) present an exemplary Voigt-fit of P14e transition in $C_2H_2$ along with the corresponding residuals. The residuals from the Voigt fit exhibit no systematic patterns, with a standard deviation of 0.59 %. We achieved a mean accuracy of 10.93 MHz, with a relative accuracy of 0.000004 % across the whole spectrum. The retrieved transition frequencies display a statistical uncertainty of 48.29 MHz. This level of precision highlights the system's ability to resolve fine details in the spectral lines, ensuring accurate detection and characterization of molecular transitions. However, incorporating a real-time frequency calibration unit, such as a hydrogen maser, could further improve the accuracy of the frequency scale. Nonetheless, the excellent agreement between the experimental and computed profiles confirms the reliability of our technique and demonstrates its suitability for precise measurements of line intensities and concentrations.

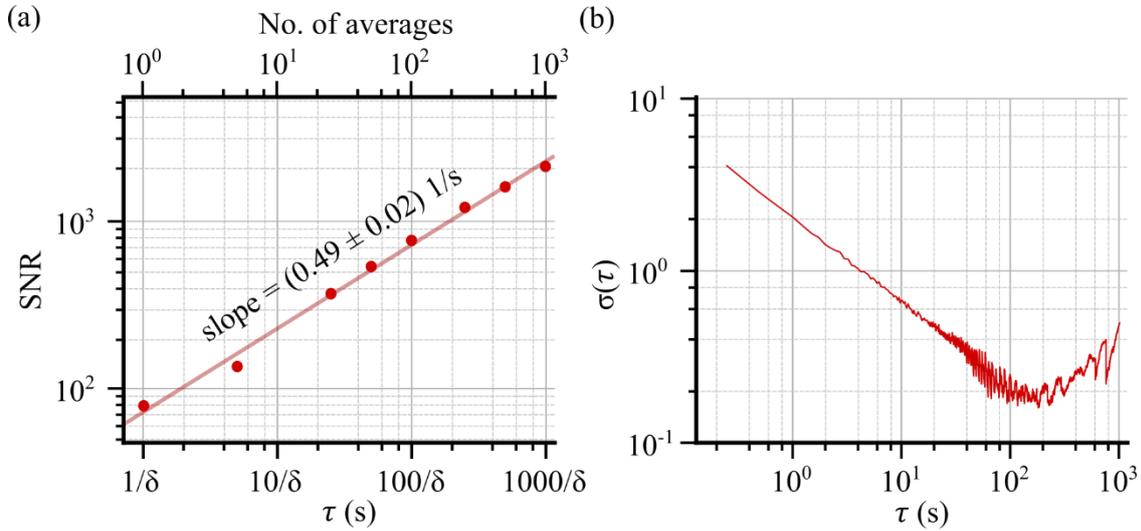

*Fig. 5. Investigation of the feed-forward stability over averaging time. (A) evolution of the signal-to-noise ratio (SNR) with measurement time. The detuning δ was set to 24 Hz. Interferograms are time-averaged and Fourier-transformed with 80 MHz resolution. A linear fit with a slope of 0.503(5) confirms the SNR scales with the square root of the measurement time, as expected for coherent averaging. (B) Allan variance of the peak frequency of the RF spectrum*

Next, we assess the stability of the phase locking by analyzing the SNR dynamics over time and by evaluating the Allan variance of the position of the Gauss-fit peak frequency in the RF spectra of the DCS interferogram. Figure 5(b) demonstrates the evolution of the SNR as a function of the square root of time (or the number of averages) near the center of the spectrum, around 288.6 THz, in the absence of an absorber. With 1,000 averages, the system achieves an SNR of 2045 for a measurement time of 41.667 seconds, corresponding to 312.69 $s^{-1/2}$. The average SNR across an 8 THz spectral bandwidth is approximately 427 ( 66.20 $s^{-1/2}$), this corresponds to a resulting figure of merit of 6.6 x$10^6$ $Hz^{1/2}$, which is comparable to that of the original dual-comb spectroscopy feed-forward methods in NIR [18] and mid-IR [26]. To determine the effective coherent averaging time, we monitored the variations in the peak frequency of the RF spectra over extended periods under locked conditions. Individual IGMs are recorded over an extended time span, and the peak frequency is determined by applying a Gaussian fit to the spectra, which are derived from the Fast Fourier Transform (FFT) of each IGM. An Allan deviation analysis revealed a minimum deviation at around 100 seconds, indicating this as the optimal averaging time. Beyond this, averaging up to 1000 seconds remains feasible, because the deviation is approximately 0.48 Hz in the RF domain and 1.6 MHz in the optical domain, which is still below the spectrometer's resolution. However, longer averaging times are currently impractical due to significant drift in the $f_{ceo}$ of both lasers. This can be improved by incorporating a slow feedback loop for individual $f_{ceo}$ stabilization.

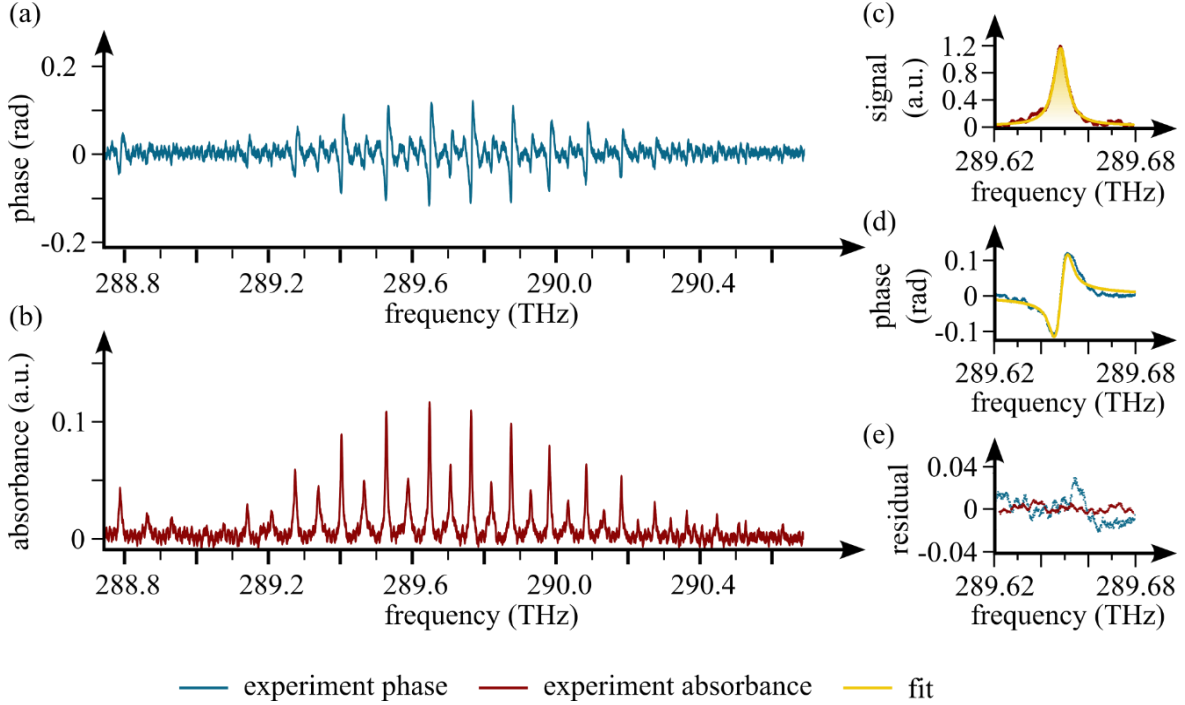

*Fig. 6. Phase information and absorbance from a dispersive DCS set-up, investigating acetylene. (A) absorbance spectrum across the measured spectral region. (B) phase information derived from the FFT signal, aligned with the positions of absorption peaks. (C) a Lorentzian fit (yellow line) applied to the strongest absorption peak at 289.65 THz. (D) typical derivative profile with Dawson function fit of the phase for the same absorption peak shown in (c). (E) the residual values between the fits and measurement values from panels (C) and (D).*

Finally, we utilize the phase-locked DCS system to measure absorbance and phase using a dispersive optical configuration. Amplitude and phase information are extracted by applying a complex Fourier transform to the coherently time-domain averaged interferograms. Figure 6(a) presents the measured frequency-domain response of the absorbance and, while Figure 6(b) phase spectra across the entire spectral region of the $C_2H_2$ ro-vibrational transitions. The absorption lines are well-separated, following Lorentzian profiles due to the high pressure in the gas cell, while the phase shift exhibits a characteristic "derivative" shape with overlapping baselines near adjacent transitions. However, both spectra exhibit an elevated noise level due to the unavoidable etalon effect inherent in the dispersive configuration. Figure 6(c) shows the absorption signal with a Lorentzian fit, and Figure 6(d) presents the corresponding dispersion signal fitted with a Dawson function. Both dispersion and absorbance signals indicate the same line positions. However, the fit for the dispersion signal is less accurate, as it is difficult to assess the baseline due to the high pressure, resulting in residual spikes except at the center of the Dawson fit. Additionally, the phase measurements show a scatter of 7.6 mrad, corresponding to a timing jitter of 4.1 attoseconds between comb lines. Although slightly higher, this value is of the same order as other coherent DCS systems [28]. Such an excellent level of phase stability in the phase-locked dual-comb spectrometer system opens the door to numerous time-domain applications, including optical waveform analysis, precise distance measurements, and dispersion characterization.

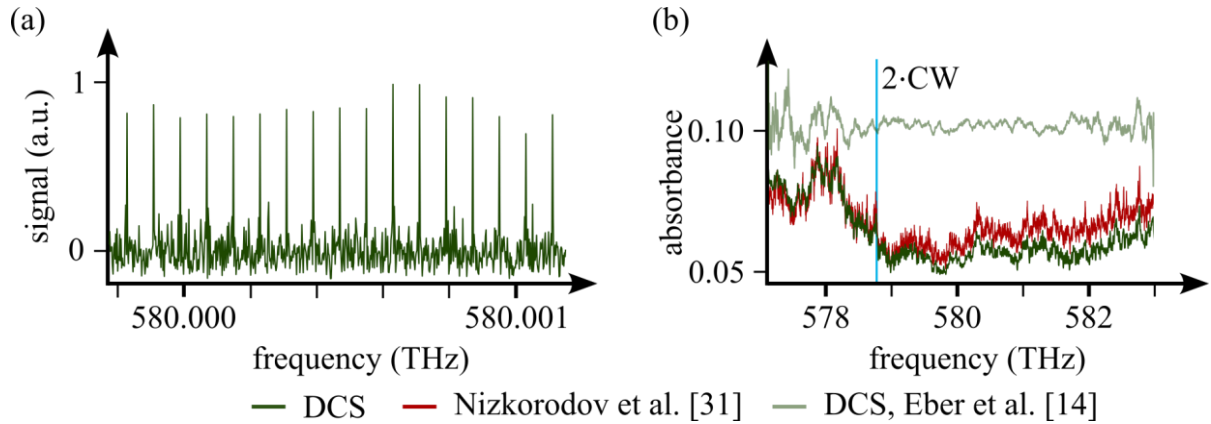

*Fig. 7. Comb-resolved spectrum and NO$_2$ absorbance analysis using visible DCS. (A) high-resolution comb-resolved DCS spectrum with a mode spacing of 80 MHz, showing distinct comb lines. (B) absorbance spectrum of NO$_2$ measured using the visible DCS setup, compared against Nizkorodov et al. [31] and our previous DCS measurements [14].*

In addition to the infrared region, we have extended our phase-stable dual-comb spectroscopy (DCS) approach to the visible wavelength range, demonstrating its capability across different spectral windows. Two visible frequency combs, centered at 517 nm, are generated through individual second harmonic generation (SHG) using two lithium triborate (LBO) crystals. The SHG process is performed after the NIR beam passes through the AOFS. Figure 7(a) presents a section of the comb-resolved visible DCS spectrum, derived from the FFT of several interferograms (IGMs), showing narrow comb modes with an 80 MHz spacing. This demonstrates the robustness of our relative phase-stabilization method, successfully preserving mutual coherence between the two frequency combs in the visible region even after a non-linear process. Notably, the AOFS driver minimizes amplitude fluctuation effects from the locking process, ensuring consistent power output in the visible comb generation. To showcase the system's real-time applicability in the visible spectral region, we investigate the absorption profile of NO$_2$ as a prototype molecule. NO$_2$ is a key trace pollutant in the atmosphere, and its precise detection is crucial for understanding atmospheric chemistry and assessing air quality [29,30]. Figure 7(b) displays the absorbance spectrum of NO$_2$, which shows excellent agreement with Nizkorodov et. al. [31] and a significant improvement over our previous DCS measurements [14] with an unlocked system. The extended mutual coherence and high resolving power of the spectrometer enable effective averaging of the NO$_2$ spectrum, significantly lowering the detection limit by a factor of 5 to 1 ppb and enhancing the selectivity for targeted gas sensing.

**Conclusion**

In this work, we develop and demonstrate a robust coherent averaging technique for a ytterbium-based dual-comb spectrometer operating at 1.03 μm wavelength, utilizing two mode-locked femtosecond lasers. Our approach effectively compensates for relative phase fluctuations between the two frequency combs by incorporating a high-speed acousto-optic element and an unstabilized continuous-wave (CW) laser source. By maintaining mutual coherence between the combs for up to 1000 seconds, we perform precise time-domain averaging of the IGMs, significantly enhancing the SNR and spectral resolution. To validate this technique, we successfully measure a comb-resolved spectrum of the $C_2H_2$ ro-vibrational transitions, specifically focusing on the weak overtone and combination bands of the symmetric C-H stretch. The high SNR (300) achieved through coherent averaging confirms the effectiveness of our technique. We further compare the measured absorption line transition properties with the HITRAN database, finding good agreement within the expected uncertainty bounds. This reinforces the precision of our measurements and the reliability of our approach.

We demonstrate the capability of multi-heterodyne spectroscopy using phase-locked frequency combs to efficiently measure both the absorption and phase spectra of molecular samples across a broad spectral range. Moreover, we extend this coherent averaging technique to the visible wavelength range, showcasing its strong potential for high-sensitivity and selective trace gas detection, even at different wavelength intervals. The success of this technique opens up new possibilities for high-sensitivity, broadband sensing applications that require both high spectral and temporal resolution. The implementation of multi-pass cells, as well as the extension of this method to shorter wavelengths capable of probing electronic transition, presents promising avenues for future research. Looking forward, the application of this coherent averaging technique in trace gas detection at visible and ultraviolet wavelengths, already feasible with ytterbium systems, has the potential to further enhance detection capabilities in environmental monitoring, medical diagnostics, and industrial processes. The stability and precision of our dual-comb spectrometer surpass conventional FTS methods, and these attributes will undoubtedly stimulate the development of advanced applications in spectroscopy, metrology, and beyond.

**Acknowledgments**

The authors would like to acknowledge Nathalie Picqué, Manish Garg and Roland Lammegger for their fruitful discussions.

**Funding:** HORIZON EUROPE European Research Council (947288); Austrian Science Fund (FWF) (Y1254). M.O. acknowledges funding from the European Union (grant agreement 101076933 EUVORAM). The views and opinions expressed are, however, those of the author(s) only and do not necessarily reflect those of the European Union or the European Research Council Executive Agency. Neither the European Union nor the granting authority can be held responsible for them.

**Author contributions:** M.P., A.E., M.O. and B.B. conceived the concept of the work and designed the system. M.P. and A.E. constructed the optical systems with the help of M.O. M.P., A.E., L.F., and E.H. performed the experiments. M.P. and A.E. analyzed the data. B.B. supervised the work. M.P. and A.E. wrote the manuscript with inputs from B.B. L.F. and E.H. were involved in the discussion

**Competing interests:** The authors declare no conflicts of interest.

**Data availability**

Data underlying the results presented in this paper are not publicly available at this time but may be obtained from the authors upon reasonable request.